\documentclass[preprint]{aastex} 

\slugcomment{Submitted to ApJ Letters}
 
\shorttitle{Raman Scattering in M2-9, RR Tel and He 2-106}
\shortauthors{Lee, Kang \& Byun}
  
\begin{document} 
   
\title{Raman Scattered \ion{He}{2} Line in the Planetary Nebula M2-9
and the Symbiotic Stars RR Telescopii and He 2-106}

\author{Hee-Won Lee$^{1,2}$, Yong-Woo Kang$^1$ \& Yong-Ik Byun$^1$}
\affil{
$^1$ Department of Astronomy, Yonsei University, 
       Seoul, Korea \\
$^2$Department of Earth Science, Sejong University,
Seoul, Korea }
\email{hwlee@galaxy.yonsei.ac.kr 
ywkang@galaxy.yonsei.ac.kr, byun@darksky.yonsei.ac.kr}

\begin{abstract}

In this Letter, we report the detection of an emission feature around
6545 \AA\ in the spectra of the bipolar planetary nebula M2-9 and the
symbiotic stars RR Telescopii and He 2-106 and propose to identify it as 
the \ion{He}{2} Raman scattered feature.
This feature was predicted by Nussbaumer, Schmid \& Vogel (1989), who
suggested that it is formed through Raman scattering by atomic hydrogen of 
\ion{He}{2} $n=6\rightarrow n=2$ photons with slightly
shorter wavelength than that of Ly$\beta$.  
The scattering cross section $\sim 10^{-20}{\rm\ cm^2}$ for this process 
implies the existence of a neutral hydrogen component with
a column density $N_{HI}\sim 10^{20}{\rm\ cm^{-2}}$ 
around the \ion{He}{2} emission regions in these objects, 
which are believed to be associated with the mass loss process in
the late stage of stellar evolution. Brief discussions on the astrophysical
implications of Raman scattering in these objects are presented. 
\end{abstract} 

\section{Introduction}

Schmid (1989) identified the broad emission features around
6830 \AA\ and 7088 \AA\ in symbiotic stars as the Raman scattered \ion{O}{6}
$\lambda\lambda$ 1032, 1038 by atomic hydrogen, where the scattering hydrogen
atom initially in the $1s$ state de-excites to the $2s$ state after
scattering. Observational evidence for Raman scattering by
atomic hydrogen includes strong polarization exhibited in the 6830 \AA\
and 7088 \AA\ features (Schmid \& Schild 1994, Harries \& Howarth 1996, 2000) 
and simultaneous detections of the features in the far UV and optical
regions (Birriel, Espey \& Schulte-Ladbeck 1998, 2000).

The proximity of the \ion{He}{2} emission lines corresponding to the 
transitions between energy levels with even principal quantum numbers
to the \ion{H}{1} resonance transitions requires much less 
\ion{H}{1} column density for Raman scattering than \ion{O}{6} 1032, 
1038 lines. Therefore,
we may expect that the Raman scattering of \ion{He}{2} may be observed 
in broader classes of objects than symbiotic stars.  
Thus far, the Raman scattering processes by atomic hydrogen have been
known to operate only in symbiotic stars with an exception of the
planetary nebula NGC~7027, in which P\'equignot et al. (1997) found the
\ion{He}{2} Raman-scattered feature around 4851 \AA\ that is 
slightly blueward of H$\beta$. 

The same \ion{He}{2} Raman-scattered feature has been reported in 
the spectrum of the symbiotic nova RR Telescopii by van Groningen (1993).
Therefore, it is also natural to expect that a stronger Raman-scattered
\ion{He}{2} line will be found around 6545 \AA,  blueward of H$\alpha$ in 
the same object and also probably in other symbiotic stars, as pointed out by
Nussbaumer, Schmid \& Vogel (1989).  According to Hyung and the colleagues, 
a number of planetary nebulae show the emission
feature around 6545 \AA, for which no proper identification has been
attempted (e.g. Hyung \& Aller 1995, Aller \& Hyung 1996, Hyung et al. 2000).

The planetary nebula M2-9 has been studied intensively especially regarding
its beautiful butterfly nebular morphology. It has been suggested that
the central object of M2-9 has a compact and hot source that may be necessary
to drive fast bipolar outflows and excite \ion{O}{3} lines (e.g. Schwarz et al.
1997).  Balick (1989) presented a
high resolution spectrum of M2-9 and reported that there exist very broad
wings around H$\alpha$ with velocity width $\sim 10^4{\rm\ km\ s^{-1}}$. The 
spectroscopic characteristics of M2-9
including rich and prominent emission lines lead him to conclude that
M2-9 resembles very much symbiotic novae such as RR Tel. 
Nussbaumer et al. (1989) pointed out that Raman scattering of Ly$\beta$
may give rise to broad wings around H$\alpha$, and according to Lee (2000)
and Lee \& Hyung(2000), the H$\alpha$ wings of the planetary nebula 
IC~4997 and many symbiotic stars are fitted very well with the template
wing profiles obtained from Raman scattering of Ly$\beta$.

In this Letter, we present our high resolution spectra of M2-9 and 
the symbiotic stars RR Tel and He 2-106, and report the detection of
Raman scattered 6545 \AA\ feature in these objects.

\section{Atomic Physics}

The slight difference in the reduced mass for \ion{H}{1} and 
\ion{He}{2} leads
to the higher energy associated with the $2s-2np$ transition in \ion{He}{2}
than the energy for the $1s-np$ transition in \ion{H}{1} by an amount of
$ \Delta E_n \simeq{3\over 4}{m_e\over m_p} E_{Ryd}(1-n^{-2})$ 
where $E_{Ryd}$ is the Rydberg energy for hydrogen.  
Therefore, for an incident line photon 
corresponding to the $2s-2np (n\ge3)$ transition in \ion{He}{2}, 
the wavelength shift of the outgoing (Raman-scattered) photon
from the Balmer series transition is given by
\begin{equation}
\Delta\lambda_o  
\simeq  5.9\left[{n^2(n^2-1)\over (n^2-4)^2}\right] {\rm\ \AA} .
\end{equation}

Therefore, for $n=3$ we get $\Delta\lambda_o \simeq 17$\,\AA\
and $\lambda_o \simeq 6546$\, \AA, which is slightly redward of
[\ion{N}{2}]$\lambda$ 6548. For $n=4$, we have $\Delta\lambda_o \simeq
9.9$\,\AA\ and the expected wavelength redward of H$\beta$ is
$\lambda_o \simeq 4851$\,\AA, at which the \ion{He}{2}
Raman scattered line was found in the spectra of the planetary nebula
NGC~7027 (P\'equignot et al. 1997) and the symbiotic star RR Tel 
(van Groningen 1993)

The more exact values of the wavelengths for the \ion{He}{2} transitions 
between levels with $n=2$ and $n=6$ are found in Table~1, where we also
show the scattering cross sections
in units of the Thomson scattering cross section $\sigma_T=
0.66\times 10^{-24}{\rm\ cm^2}$. The dominant transitions
are $6p_{1/2,3/2}\rightarrow 2s_{1/2}$, for which the Raman-scattering lines
are expected around 6545\,\AA\ and the Rayleigh and
Raman scattering cross sections are $\sim 7000\sigma_T$ and $\sim 1000
\sigma_T$ respectively (e.g. Lee \& Lee 1997, Nussbaumer et al. 
1989).  The \ion{He}{2} line photons with the final state 
$2p_{3/2}$ will have the Raman scattered wavelength around 6547.4\,\AA,
which will make a minor contribution to the formation of the Raman
scattered feature but will be blended with the forbidden line
[\ion{N}{2}]$\lambda$ 6548. 

The total scattering cross section,
which is the sum of the Raman and Rayleigh scattering cross sections 
for these photons, is found to be 
$\sim 8000\ \sigma_T\sim 5\times 10^{-21}
{\rm\ cm^2}$. Therefore, in the presence of a neutral component
with the column density $N_{HI}\ga 10^{20}{\rm cm^{-2}}$ around the \ion{He}{2}
emission region we may expect significant Raman scattering of
\ion{He}{2} line photons. 

\section{Observation}

We observed during the nights of August 14-16, 2000 at the Cerro Tololo
Inter-American Observatory using the Bench Mounted Echelle spectrograph
installed on the 1.5~m telescope. The exposure time was 2400 seconds for 
RR~Tel and 3600 seconds for M2-9 and He2-106 and the seeing was $\la 1^{''}$. 
Standard reduction procedures have been performed using the echelle 
spectroscopic data reduction tasks included in the IRAF package.

In Fig.~1, we show the spectra around
H$\alpha$ for RR Tel, M2-9 and He2-106. The vertical bars mark
the emission features at $\lambda=6527$\,\AA\ and 6545 \AA.
All these objects show very broad wings around H$\alpha$, where
the origin was proposed to be Raman scattering of Ly$\beta$ by 
Nussbaumer et al. (1989).
The dashed lines represent the $\Delta\lambda^{-2}=(\lambda-
\lambda_{H\alpha})^{-2}$ profile, which is the Raman-scattered profile
of the flat incident Ly$\beta$ profile (Lee 2000, Lee \& Hyung 2000).
However, we note that the
wing formation mechanism is not a fully resolved matter and that the broad 
wings may also be explained by other mechanisms including electron
scattering and fast outflows around the central star (e.g. 
Arrieta \& Torres-Peimbert 2000, Schwank, Schmutz \& Nussbaumer 1997,
van de Steene, Wood, \& van Hoof 2000).

RR~Tel and He2-106 show strong \ion{He}{2}$\lambda$ 6560 emission line
in our spectra, whereas in M2-9 it appears absent.
The H$\alpha$ emission line of RR~Tel and He2-106 shows a single peak profile
but M2-9 shows a clear double peak profile in H$\alpha$, which is 
prevalently seen in many other symbiotic stars
(e.g. Ivison, Bode, \& Meaburn 1994, van Winckel, Duerbeck, \& Schwarz 1993). 
The double peak profile (or centrally reversed profile) of the H$\alpha$
in symbiotic stars and planetary nebulae is not completely understood
(e.g. Schmutz et al. 1994, Robinson et al. 1994).

The emission line around 
6527 \AA\ in  RR~Tel and He2-106 corresponds to
the transition between $n=14$ and $n=5$ states of \ion{He}{2} and is marked by 
a vertical bar in Fig.~1.  The same feature is absent in M2-9. 
From Gaussian fittings to the 6527 and 6560 features in RR Tel
we obtain the full width at half maximum values of 1.4 \AA\ for both features.  

In order to separate the 6545 feature from the spectra, we use the fact that
the [\ion{N}{2}]$\lambda$ 6548 profile should be identical with that
of the [\ion{N}{2}]$\lambda$ 6584 feature, which is 3 times stronger
due to the atomic structure (e.g. Osterbrock 1989, Storey \& Zeippen 2000).
In Fig.~2, based on our fitting of the H$\alpha$ wings 
we isolated the emission fluxes around 6548 \AA\ and 6584 \AA.
Subsequently we subtracted the flux of the [\ion{N}{2}]$\lambda$ 6584
multiplied by 1/3 (represented by red lines) from the flux around the 
6545 and [\ion{N}{2}]$\lambda$ 6548 features (represented by the black lines)
to isolate the \ion{He}{2} Raman-scattered feature.
The result is shown by the blue lines, which clearly show the 6545 features
in these objects. We also see an additional broad emission
feature around 6581 \AA, which causes a slight excess in the 
[\ion{N}{2}]$\lambda$6584 feature.

It is apparent that the width of the 6545 \AA\ feature is much 
wider than 1.4 \AA\ of these direct \ion{He}{2} emission lines, which is 
naturally explained by the enhancement of the Doppler width by a factor 
of $\sim 6.4$ attributed to the scattering incoherency. The 6545 feature 
in RR~Tel appears to exhibit double or multiple peak profiles 
whereas single peak profiles are seen in the \ion{He}{2} emission lines 
at 6527 \AA\ and 6560 \AA. Further investigation 
using larger telescopes will shed more light on this matter.

\section{Discussion}

Since many symbiotic stars exhibit bipolar nebular morphology and M2-9
is bipolar, it is highly plausible that the neutral hydrogen component
responsible for Raman scattering is significantly aspherical.
If this is the case, then the 6545 \AA\ feature is expected to be polarized
either parallel or perpendicular to the symmetry axis. Therefore, as in
the case of the Raman-scattered \ion{O}{6} 6830 \AA\ and 7088 \AA\ features,
spectropolarimetry using large telescopes will be important 
to verify the scattering nature of the 6545 \AA\ feature.

The typical width $\sim 20$\,\AA\ of the Raman scattered 
\ion{O}{6} features around 6830 \AA\ and 7088 \AA\ in many symbiotic
stars implies that the \ion{O}{6} emission region
shows a velocity scale up to $100{\rm\ km\ s^{-1}}$. However,
the Raman scattered \ion{He}{2} line features seem to show narrower widths
($\la 10$\,\AA) than the above Raman scattered features.  
Because \ion{O}{6} 1032, 1038 lines have smaller scattering cross sections
by two orders of magnitude than \ion{He}{2} emission lines, 
it is expected that the
6545 \AA\ feature is formed in a more extended region than 6830 \AA\ and 
7088 \AA\ features, which may result in difference in the profiles.
Dumm et al. (1999) presented a broad range of neutral hydrogen column 
densities $10^{21-25}{\rm\ cm{^-2}}$ in the circumstellar region 
in the eclipsing symbiotic star SY Muscae. Various Raman scattering lines
with a large range of cross sections will play an important role to
diagnose the circumstellar region that is closely related with the
mass loss process in the late stage of stellar evolution.

Selvelli et al. (2000) performed high resolution spectroscopy of RR Tel
to reveal prominent emission lines and the broad H$\alpha$ wings,
which are consistent with our observation. However, in their observation
the H$\alpha$ red wing appears to extend up to 6300 \AA. If electron
scattering is the origin of these broad wings, then the scattering
plasma is characterized by the electron temperature $T_e\sim 10^6{\rm\
K}$ and a high free electron column density $N_e\ga 10^{23}{\rm\ cm^{-2}}$,
which appears unnatural.  Assuming that the Raman scattering of Ly$\beta$ is
responsible for these wings, because the Ly$\beta$ {\it line emission} 
is not likely to show a flat profile with this huge width, it may be more 
plausible that the broad H$\alpha$ wings are formed from the Raman scattering
of the {\it continuum} photons around Ly$\beta$. This hypothesis 
may explain the difference in the widths of the \ion{He}{2} Raman 
scattered features and the \ion{O}{6} Raman features around 6830 \AA\
and 7088 \AA. We note that all three objects (M2-9, RR~Tel, \& He2-106) 
exhibit similar broad H$\alpha$ wings, which strengthens the
hypothesis of the Raman scattering origin of the wings.

Due to the absence of 
\ion{He}{2} emission lines, M2-9 has been classified into low excitation
planetary nebula (e.g. Gurzadyan, 1997). The existence of the
6545 feature implies that 
the \ion{He}{2} emission region is highly obscured from our line of sight
whereas the scattering region is much more extended than the emission region.
This is very plausible considering that the \ion{He}{2} emission region 
is located near the central star responsible for the photoionization,
which can be hidden by an optically thick component such as a molecular
torus and is consistent with the IR observation of a disk-like
structure across the core region in M2-9 by Aspin, Maclean \& Smith (1988). 

Hyung and his colleagues showed that a number of planetary 
nebulae exhibit the emission features around 6545 \AA\ and also 6581 \AA\
(e.g. NGC 7009 by Hyung \& Aller 1995, NGC 6790 by Aller \& Hyung 1996, and
NGC 6543 by Hyung et al. 2000). These planetary nebulae also
exhibit emission features around 4851 \AA\ and 4332 \AA, which they 
attributed to \ion{Mg}{2} and \ion{O}{2}. However, as in RR Tel, 
if the 6545 \AA\ feature is the \ion{He}{2} Raman scattered line, 
then the  4851 \AA\ and 4332 \AA\ features in these planetary nebulae 
may be more adequately interpreted to be the Raman scattered \ion{He}{2} 
features.  Higher quality spectroscopy using 
large telescopes is needed to obtain more reliable profiles of these 
features for comparison with other direct emission lines.

\acknowledgements
We thank the staff at CTIO for their support in telescope
operation. We also thank Dr. Siek Hyung, who sent us his spectra 
secured from the Hamilton echelle spectrograph mounted at the Lick 3m 
telescope.  This work was supported by Korea Research Foundation 
Grant KRF-1999-115-DP0441.

\figcaption[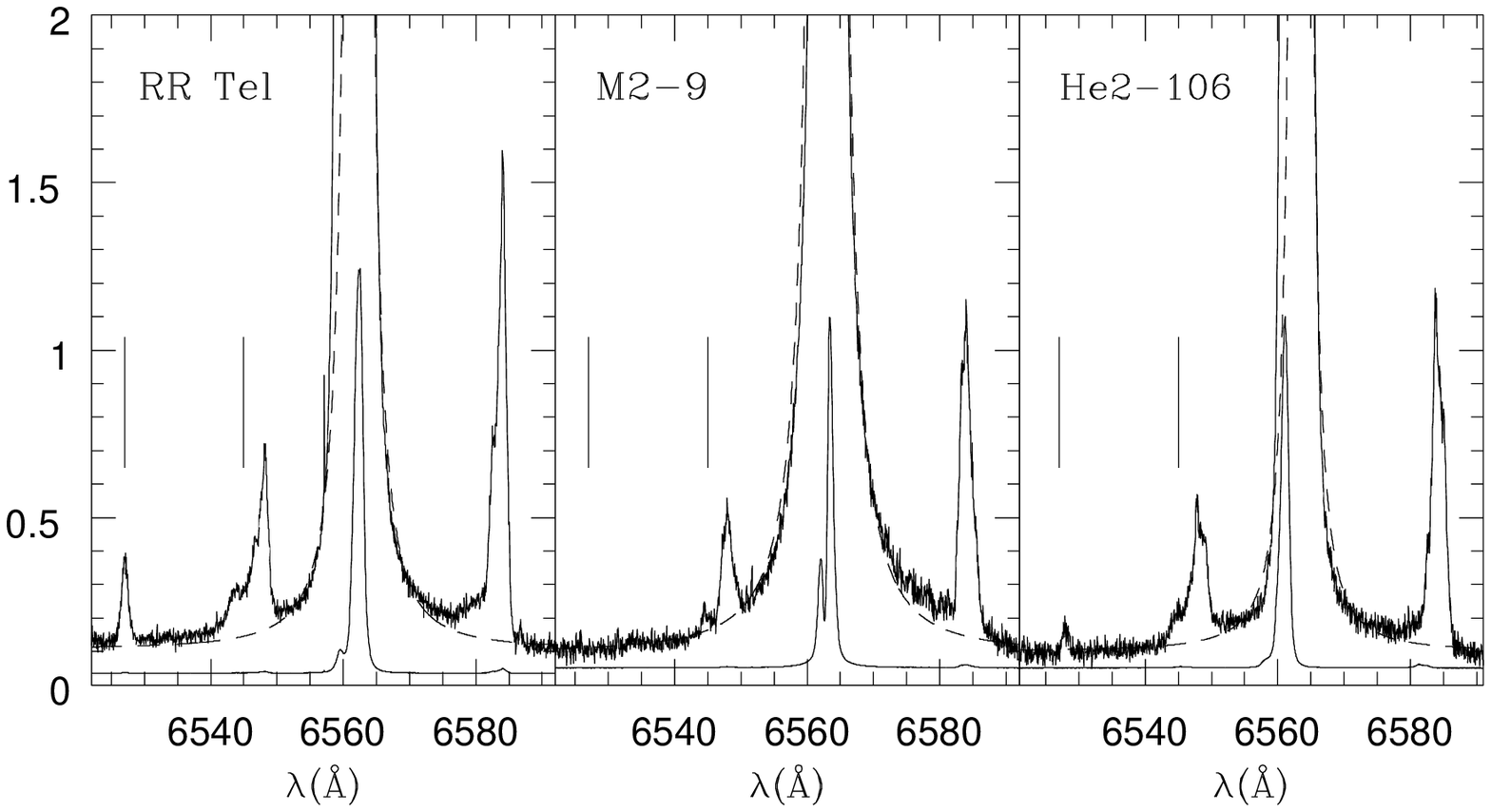]{
The spectra around H$\alpha$ of RR Tel, M2-9 and He2-106 obtained from
the Bench Mounted Echelle Spectrograph mounted on the 1.5 m telescope
at the Cerro Tololo Inter-American Observatory. We illustrate the flux 
multiplied by 1/100 to show the whole H$\alpha$ profile.
The vertical bars mark
the emission features at $\lambda=6527{\rm\ \AA}$ and
6545 \AA, where the former arises from the transition between 
$n=14$ and $n=5$ states of \ion{He}{2} and the latter is proposed to be
the \ion{He}{2} Raman-scattered feature by atomic hydrogen. All the objects 
show similar broad
H$\alpha$ wings that are excellently fitted by the $\Delta\lambda^{-2}
=(\lambda-\lambda_{H\alpha})^{-2}$ profile expected for the flat
Ly$\beta$ flux Raman-scattered by neutral hydrogen and
represented by the dashed lines.
}
\figcaption[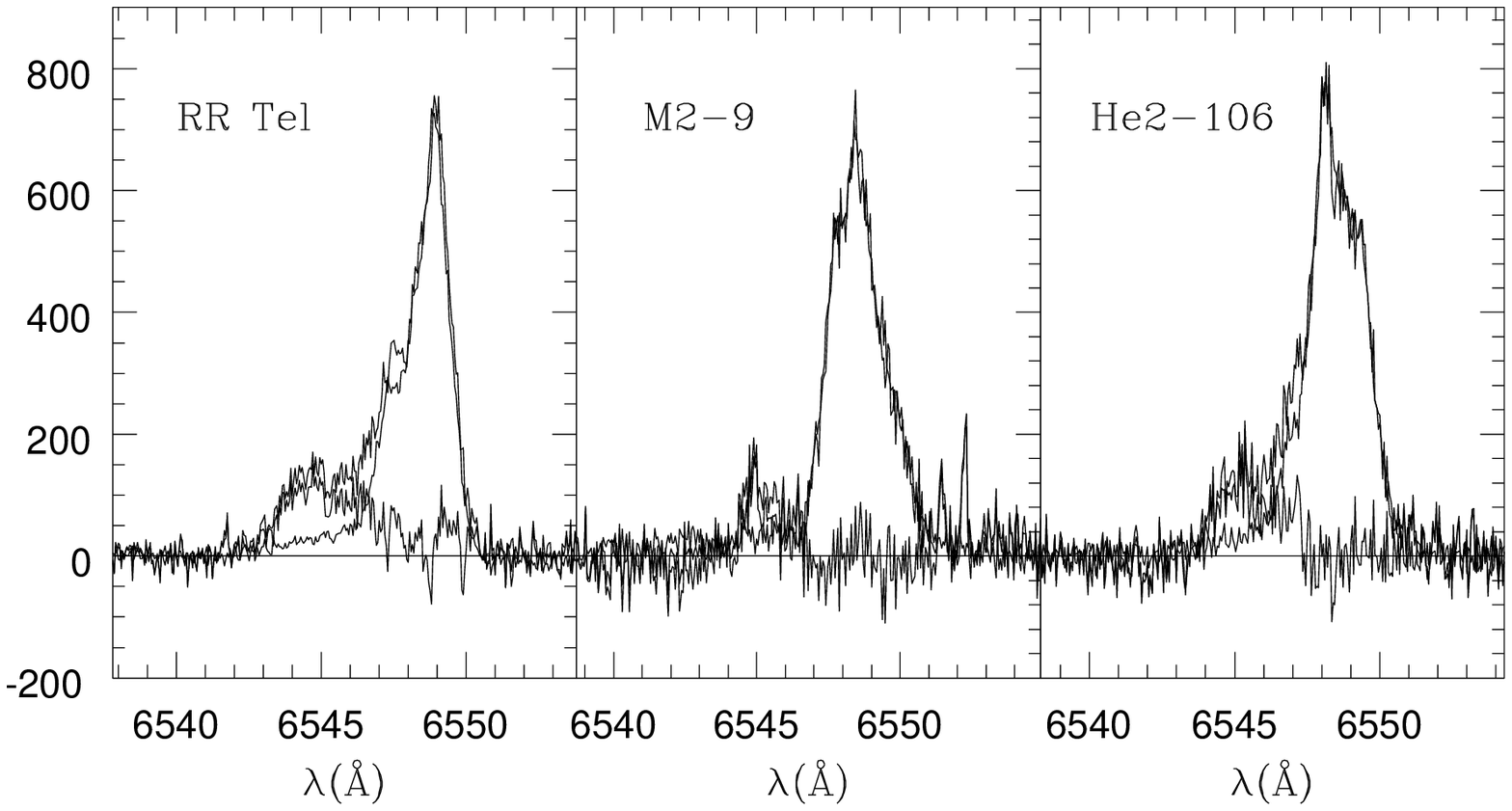]{
The \ion{He}{2} Raman scattered feature around 6545 \AA. 
Based on our fitting of the H$\alpha$ wings,
we subtracted out the H$\alpha$ wing flux to isolate the emission fluxes 
around 6548 \AA\ and 6584 \AA. The black lines
represent the flux around the 6545 feature and the [\ion{N}{2}]$\lambda$6548
emission line.  We plot by the red lines the flux 
around [\ion{N}{2}]$\lambda$6584 multiplied by 1/3 and subtract it from the 
[\ion{N}{2}]$\lambda$6548. The remainder is shown at the bottom in each
panel by the blue solid lines, from which the 6545 feature is clearly seen.
}

\begin{deluxetable}{ccccc}
\footnotesize
\tablecaption{Wavelengths for the transitions of \ion{He}{2} between $n=2$
and $n=6$ states} 
\tablewidth{0pt}
\tablehead{
\colhead{ Transition }&
\colhead{ Wavelength (\AA) }  &
\colhead{ $\lambda_{Ram}^a$ (\AA) } &
\colhead{ $\sigma_{Ram}^b$ } &
\colhead{ $\sigma_{Ray}^c$ }
}
\startdata
$6s_{1/2}\rightarrow 2p_{1/2}$ & 1025.24309 & 6545.3782 & 938
&7257\\
$6s_{1/2}\rightarrow 2p_{3/2}$ & 1025.30452 & 6547.4605 & 1238
&9530\\
$6p_{1/2}\rightarrow 2s_{1/2}$ & 1025.24821 & 6545.1820 & 958
&7412\\
$6p_{3/2}\rightarrow 2s_{1/2}$ & 1025.24593 & 6545.0890 & 949
&7344\\
$6d_{3/2}\rightarrow 2p_{1/2}$ & 1025.24029 & 6544.8568 & 927
&7173\\
$6d_{3/2}\rightarrow 2p_{3/2}$ & 1025.30241 & 6547.3915 & 1225
&9431\\
$6d_{5/2}\rightarrow 2p_{3/2}$ & 1025.30166 & 6547.3607 & 1220
&9398\\
\enddata
\tablenotetext{a}{Wavelength of the Raman-Scattered Photon}
\tablenotetext{b}{Raman scattering cross section in units of the Thomson 
scattering cross section $\sigma_T = 0.66\times 10^{-24}{\rm\ cm^2}$.} 
\tablenotetext{c}{Rayleigh scattering cross section in units of 
$\sigma_T$.} 
\end{deluxetable}
\end{document}